# Templet: a markup language for concurrent programming

## Sergey V. Vostokin[1]

*Dept. of Information Systems and Technology,*
*Samara State Aerospace University named after S.P. Korolyov,*

*Moskovskoye Shosse 34, 443086, Samara, Russian Federation*

## SUMMARY

In this paper we propose a new approach to the description of a network of interacting processes in a traditional programming language. Special programming languages or extensions to sequential languages are usually designed to express the semantics of concurrent execution. Using libraries in C++, Java, C#, and other languages is more practical way of concurrent programming. However, this method leads to an increase in workload of a manual coding. Besides, stock compilers can not detect semantic errors related to the programming model in such libraries. The new markup language and a special technique of automatic programming based on the marked code can solve these problems. The article provides a detailed specification of the markup language without discussing its implementation details. The language is used for programming of current and prospective multi-core and many-core systems.

## 1. INTRODUCTION

The Templet language is a domain-specific markup language. It is designed to be used together with a sequential procedural or an object-oriented programming language. The new property of the language is an explicit specification of the process-channel computation semantics with only marked serial code used.

The report is written as the initial specification for the markup language. It is a basis for its implementation and further documentation. Some details were intentionally omitted because they can be derived from the given code samples or they can unnecessarily restrict the freedom of implementers.

The design concepts of the language basically follow the concept of the language-oriented programming [1,2]. The algebraic-like notation similar to the CSP formalism was applied to describe processes and interactions [3]. The idea of a minimalistic design with emphasis on the basic abstractions is taken from the programming language Oberon [4].

The markup language is an application development tool for multiprocessor parallel systems. The experimental preprocessor and code samples in marked C++ are available at http://the-markup-language-templet.googlecode.com. Currently we are using the language in scientific computing service http://templet.ssau.ru.

## 2. SYNTAX AND VOCABULARY

A language is an infinite set of sentences. The sentences are well formed according to language syntax. In the Templet language these sentences are called *modules*. The module is a single file or a group of logically related files. In addition, the content of files also belongs to a programming language, which is called the *base language*. The base language is a carrier for an execution

---

[1] Correspondence to: Sergey Vostokin, E-mail: sergey.vostokin@gmail.com.

semantics of the markup language, namely communicating sequential processes model (see Sec. 4.3).

To describe the syntax, an extended Backus-Naur Formalism called EBNF is used. Square brackets [ and ] denote optionality of the enclosed sentential form. Braces { and } denote repletion (possibly zero times). Parentheses ( and ) are used for grouping of the symbols in right side of syntactic rules. Left and right sides are separated by = sign. A dot marks the end of rules.

Non-terminal symbols (which can be on the left side of syntax rules) are denoted by English words expressing their intuitive meaning.

Symbols of the language vocabulary (terminal symbols) are block symbols, delimiters, and identifiers. *Block symbols* and their corresponding lexical procedures are described in Sec. 3.

*Delimiters* are single characters from the set `'~ = ; . + ? ! | , * : & ( ) < >'` enclosed in quotation marks and the pair of characters `'->'` enclosed in quotation marks. Delimiters are located in the scheme of a module (see. Sec. 3).

*Identifiers*, denoted by `ident` symbol, are sequence of letters, digits, and other characters containing no white-spaces and delimiters. The identifiers are also located in module scheme.

## 3. CODE STRUCTURE

The following EBNF rules describe the block structure of a module.

```
module = {base-language|user-block} module-scheme {base-language|user-block}.
user-block = user-prefix base-language user-postfix.
module-scheme = scheme-prefix { channel | process } scheme-postfix.
```

Module code consists of a single *module scheme* section and multiple code sections in the base language with highlighted user blocks. These sections are distinguished from the rest of the code by means of comments provided in the base language. For example, the marked C++ [5] code may look as follows. The blocks' names according to the markup language syntax are shown on the right side.

```
#include <runtime.h>              <-- base-language

/*templet$$include*/             <-- user-prefix
  #include <iostream>            <-- base-language
/*end*/                          <-- user-postfix

/*templet*                       <-- scheme-prefix
  *hello<function>.              <-- module-scheme
*end*/                           <-- scheme-postfix

void hello(){                    <-- base-language
/*templet$hello$*/               <-- user-prefix
  std::cout << "hello world!!!"; <-- base-language
/*end*/                          <-- user-postfix
}                                <-- base-language
```

Lexical analyzer defines the boundaries of the blocks by signatures, recognizing specific sub-strings in a character stream. For example, the module scheme may be preceded by a combination of characters `/*templet*`, and finish by `*end*/`. User block prefixes include identifiers for binding the blocks with module scheme: `/*templet$hello$*/` bound with `*hello<function>. .`

The module is a program skeleton, and user blocks are extension points. Module scheme defines the structure of program skeleton.

Markup language provides a *mapping algorithm*. Mapping is a module transformation carried out by rewriting the module code. Mapping is applied only to a module with syntactically correct scheme. As a result of this transformation the code and the scheme becomes isomorphic meaning that the scheme can be reproduced from the code and vice versa. New user blocks may appear. Existing user blocks may move to new positions or turn into comments.

# 4. CLASSES

A module scheme includes definitions of *classes*: channels and processes. A *channel* scheme has the form `~ name [body]. `. A *process* scheme has the form `* name [body]. `. Channels describe communication of processes that perform calculations.

For example, supposedly there is a program that checks the trigonometric identity $\sin^2 x + \cos^2 x = 1$. When process of `*Master.` class sends $x$ values to working processes of `*Worker.` class via channels of `~Link.` class. The master gets the squares of trigonometric functions and calculates their sum in return.

Class is identified by its `name`. The name is unique within a module scheme, and visible after the point of its definition. Class `body` also has identifiers defined within it. Such identifiers are visible at any point inside the body of the class.

The mapping algorithm should ensure adherence to the sequencing rule for classes: class code appears in the same order as it was for class scheme. This is necessary for the programming languages in which the class declaration should be preceded by a reference to it.

## 4.1. Parameters

With the *parameters* one can define a specific algorithm for mapping class schemes into the base language code.

```
params = '<' ident {',' ident} '>'.
```

For instance, if the verification of trigonometric identity uses a request-response protocol, the channel can be defined as `~Link <request-response>. ` If the verification of trigonometric identity uses master-worker pattern, the pattern can be defined as a single process `*Pythagorean-identity-test <master-worker, shared-memory>. `

## 4.2. Rules and Messages

The *rules* define processing of *messages* in channels and processes.

```
rules = rule { '|' rule }.
rule = ident { ',' ident } '->' ident.
```

Rules are defined as the set of pairs `message -> state.` Each pair denotes a message and its corresponding transition state. Symbol '|' separates one rule from another. Abbreviated notation `A,B -> S` can replace `A->S | B->S` in case of repeating states.

## 4.3. Classes Mapping

Code of the classes presents a programming model semantics for the markup language. Language class (channel or process) inherits its behavior from run-time class `BaseChannel` (for a channel) or `BaseProcess` (for a process). The language does not define specific interface or implementation for `BaseChannel` and `BaseProcess`. But they should be implemented in such a way that the following behavior of `Channel` and `Process` classes is possible.

```
class Channel: public BaseChannel{
 public:
  // test whether the channel it accessible
  bool access_client(){...} // at client side
  bool access_server(){...} // at server side
  // client sends entire channel to server
  void send_client(){...}
  // server sends entire channel to client
  void send_server(){...}
...
};
class Process: public BaseProcess{
 public:
  // receive data on the channel
  virtual void recv(BaseChannel*);
  // bind a channel to the process as client
  bool bind_client(BaseChannel*){...}
  // .. or server
  bool bind_server(BaseChannel*){...}
...
}
```

Class `BaseChannel` should be such that it is possible to implemente the `Channel` class with the following properties. Access to the channel alternately belongs to pair of processes called client and server. Client has access right to the channel in the beginning of computations. Methods `access_client()` and `access_server()` allow client or server process to check for access. Methods `send_client()` and `send_server()` can be used to grant access from client to server process or from server to client process respectively.

Class `BaseProcess` should be such that it is possible to implemente the `Process` class with the following properties. Methods `bind_client()` and `bind_server()` establish connection of the process with the given channel as a client or as a server. Method `recv()` is called at the moment getting access to the channel. The channel is passed as an argument. If the process gets access to multiple channels, it takes several consecutive calls to `recv()` in random order. If some process sends access to channel to another process, the other process will sooner or later get access to this channel.

## 5. CHANNEL

The channel scheme defines the communication protocol between client process and server process.

```
channel = '~' ident [params] ['=' state {';' state}] '.'.
```

The protocol is specified as a set of rules. Each rule denotes a changing of *states* under the influence of messages passing through the channel.

### 5.1. State

State has a unique identifier within the channel scope. State definition consists of attributes and a set of transition rules.

```
state = ['+'] ident [ ('?'|'!')  [rules] ].
```

Plus sign `'+'` is an attribute of initial state. Question mark `'?'` denotes *question-messsage* that passed from client to server. Exclamation mark `'!'` denotes *answer-message* that passed form server to client.

For example, a request-response protocol can be defined in the way as shown below.

```
~Link = +BEGIN? Request -> PROCESSING;
         PROCESSING! Response -> BEGIN.
```

Channel protocol to verify the trigonometric identity may look like the following code.

```
~Link = +BEGIN ? ArgCos -> CALCCOS | ArgSin -> CALCSIN;
         CALCCOS ! Cos2 -> END; CALCSIN ! Sin2 -> END.
```

It is permitted to omit definition for a state from which there are no transition rules. The state with no transition rules and attributes (`'?'` or `'!'`) is considered as a client state (with `'?'` mark used).

## 5.2. Channel Mapping

Channel scheme mapping corresponds to the following code in the base programming language. This code is an interface for an access to channel in the context of a process message handler `recv()`.

```
class Channel : public BaseChannel{
...
public:
  struct Message{// message data structure
/*templet$Channel$Message*/
  // user block
  ...
/*end*/
  };

  // class field for storing the Message
  Message Message_get;

  // access tests for reading-writing the Message
  bool Message_read_client(){...}// on server
  bool Message_write_client(){...}
  bool Message_read_server(){...}// or client side
  bool Message_write_server(){...}

  // send the Message from server to client
  void Message_send_server(){...}
  // send the Message from client to server
  void Message_send_client(){...}
...
};
```

Such fragments are repeated for each message in the channel definition.

# 6. PROCESS

The process scheme defines an algorithm for processing incoming messages and sending response messages.

```
process = '*' ident [params] ['=' ((ports [';' actions]) | actions) ] '.'.
```

The process scheme includes definitions for *ports* and *actions*. Also it has the *calls* for *user functions* during the actions.

## 6.1. Ports

The port denotes channel-to-process binding point. The port is unique within a process to which it belongs. The port definition consists of attributes and transition rules for actions (that run when channel messages appear on the port).

```
ports = port {';' port}.
port  = ident ':' ident ('?'|'!')[(rules ['|' '->' ident])|( '->' ident)].
```

Port identifier is followed by colon `':'`, and *port channel* attribute is specified after the colon. Question mark `'?'` is an attribute of the *server-port*. Exclamation mark `'!'` is an attribute of the *client-port*. Messages in transition rule that follow question mark `'?'` are considered as *question-messages*, whereas messages that follow exclamation mark `'!'` are considered as *answer-messages*.

For example, the choice of the square of sine or the square of cosine calculation in trigonometric identity test depends on particular message: `p:Link ? ArgSin -> sin2 | ArgCos -> cos2` . Abbreviated form `p : Link ? -> sin2` denotes that the action `sin2` runs for all implied messages.

## 6.2. Port Mapping

The port mapping corresponds to the following code in the base programming language. The code fragment shows a part of message handler procedure `void recv(BaseChannel*c)` for the port `Port:Channel!Message1->Action1|Message2->Action2|->SomeAction` .

```
class Process : public BaseProcess{
public:
  // bind Port to channels of type Channel
  bool Port_bind_client(Channel*){...}

  // optional Port's user procedure
  void Port_call(Channel*){
/*templet$Process$Port*/
  // user block
...
/*end*/
  }
  // port field for storing the binding to a channel
  Channel* Port_port;

  // part of message handler procedure for Port in recv()
  void recv(BaseChannel* c){
    int sel;//selector that stores port tag
    sel = c->selector;//that comes from channels

    switch(sel){
      case Port_label:
      {// static cast to the Channel type
        Channel* _c=static_cast<Channel*>(c);

        Port_call(_c);  // optional call to the port procedure
        Port_port = _c; // re-assign port in case
                        // when many channels are bound to one port
        sel=UNKNOWN;

        // test for possible messages
        if(_c->Message1_read_client()) then sel=Action1_label;
        else if(_c->Message2_read_client())then sel=Action2_label;
        else sel=SomeAction_label;
```

```
      assert(sel!=UNKNOWN);// abort execution if
      break;                // unknown message has come
    }//case
    ...
  }//switch
 }//recv
};
```

The code is repeated for each port in the process. If a port was defined as server-port (with a question mark `'?'`), methods in the code will have suffices `_server` instead of `_client`.

## 6.3. User Function

User function represents a call to user-defined code in a process. The call has a name and a list of arguments.

```
call = ident '(' [args] ')'.
args = ident ('?'|'!') ident {',' ident ('?'|'!') ident}.
```

Argument list consists of comma separated port-message pairs. If the name of the port is followed by question mark `'?'`, the message is read. If the name of the port is followed by exclamation mark `'!'`, the message is written for further sending.

The method that implements user-defined calculations should return 'true' or 'false' according to the result of message processing.

## 6.4. Actions

Action specifies a sequence of user function calls and a sending of messages. It is unique within the process scope to which it belongs. The action definition consists of attributes and one or many user functions.

```
actions = action {';' action}.
action  = ['+'] [ident ':'] disjunction ['->' ([ident] '|' ident) | ident].

disjunction = conjunction { '|' conjunction}.
conjunction = call {'&' call}.
```

Plus attribute `'+'` indicates the action that is triggered at the beginning of calculations. Label before comma sign `':'` can be used as a unique action identifier. If the label is omitted, the action identifier is the name of the first user function of the action. The action to be performed in case of a successful completion of current action can be specified after the sign `'->'`. Action that runs on unsuccessful completion of current action can be specified after `'->'` `'|'` signs.

For example, processes for checking the trigonometric identity can be defined as follows.

```
*Master =
      p1:Link ! Sin2 -> join; p2:Link ! Cos2 -> join;
     +fork(p1!ArgSin,p2!ArgCos); join(p1?Sin2,p2?Cos2).

*Worker =
      p : Link ? ArgSin -> sin2 | ArgCos -> cos2;
     sin2(p?ArgSin,p!Sin2); cos2(p?ArgCos,p!Cos2).
```

By applying attributes to construct a chain of actions, `Worker` process can be defined as follows.

```
*Worker =
     p : Link ? -> DO;
   DO:sin2(p?ArgSin,p!Sin2)->|cos2; cos2 (p?ArgCos,p!Cos2).
```

The user function calls can be separated by conjunction '&' and disjunction '|' signs. The sing '&' has a priority over '|' sign. Grouping of calls is performed from left to right. The short-circuit evaluation semantics is used in calls to multiple user functions. This means that action `A()&B()->C|D` is equivalent to actions `A()->B|D; B()->C|D`. And action `A()|B()->C|D` is equivalent to actions `A()->C|B; B()->C|D`.

It is possible to define the `Worker` process with grouping of user functions in a single action.

```
*Worker =
     p : Link ? -> DO;
   DO:sin2(p?ArgSin,p!Sin2)|cos2(p?ArgCos,p!Cos2).
```

## 6.5. Action Mapping

Action mapping corresponds to the following code in the base programming language. The code fragment shows a part of message handler procedure `void recv(BaseChannel*c)` for the action `Action(p1?Message1,p2?Message2)->A1|A2`. It is assumed that ports are defined as `p1:Channel!` and `p2:Channel?`.

```
class Process : public BaseProcess{
public:
   // set function for the Action
   bool Action_call(Channel::Message1*m1,Channel::Message2*m2){
/*templet$Process$Action*/
   // user block
/*end*/
   }
  // part of message handler procedure for Action in recv()
  void recv(BaseChannel* c){
    int  sel; // selector
    bool res; // used in actions processing
        ...
    // first action in the loop was selected in port section (see Sec.6.2)
    for(;;)switch(sel){
      case Action_label:
      { //test the access for reading-writing messages
        res=p1_port->Message1_read_client()&&p2_port->Message2_write_server();

        //if test is OK, call user function
        if(res)res=Action_call(&p1_port->Message1_get,&p2_port->Message2_get);

        //if previous steps are OK, send message(s)
        if(res){p2_port->Message2_send_server();}

        //at last if all steps passed successfully, go to the action A1,
        //otherwise go to the action A2
        if(res) sel=A1_label; else sel=A2_label;
        break; // go to the next loop
      }//case
          ...
    }//for loop
  }//recv
};
```

The code is repeated for each action in the process. Complex actions comprising two or many user functions are converted into regular form `A()->B|C` beforehand.

The process implementation in base language should ensure adherence to the following calculation rules for actions. User function activates when (1) control is passed to its action; (2) it is possible to read (`'?'` mark) or write (`'!'` mark) in all the messages in its argument list. Message (marked with `'!'`) are sent when user function was activated and returns 'true'. If user function was activated and returns 'true', control goes to label `-> A`. If function was not activated or returns 'false', control goes to label `-> |A`. If no labels exist, the handler procedure ends.

# 7. PROGRAM

A program is a *network of objects*. Objects are instances of classes in the base programming language. These classes are in turn derived from channel or process schemes.

## 7.1. Object Network

The network of objects is defined in the base programming language. For example, trigonometric identity test has the following network of objects.

```
void main(){

  TempletProgram p;        // program execution engine

  Link link1(p),link2(p);// channel objects
  Master m;                // process objects
  Worker w1,w2;

  // bind channels to main Master process
  m.p1_port(link1);m.p2_port(link2);

  // bind channels to Worker processes
  w1.p_port(link1);w2.p_port(link2);

  // object network is ready, input x value
  cout<<"input x:"; cin>>m.x;
  // and run it
  p.run();
  cout<<"sin2(x) + cos2(x) = "<<m.sin2x_plus_cos2x;
}
```

Any program in markup language Templet is a network data structure of arbitrary complexity.

## 7.2. Execution

Program implementation in the base programming language should provide the opportunity for non-deterministic performance, which follows from programming model definition in Sec. 4.3. Non-determinism of program execution is simulated by means of pseudo-random numbers.

```
void TempletProgram::run()
{
  size_t rsize;
  // while message queue is not empty
  while(rsize=ready.size()){
    //select random channel which is currently sending message
    //then exclude this channel from the message queue
    //and move it to not sending state
    int n=rand()%rsize; auto it=ready.begin()+n;
    BaseChannel*c=*it;  ready.erase(it); c->sending=false;
    //extract the process to which the message was sent from the channel
    //run message handling method recv() for the channel and
```

```
    //pass the channel as the argument to this method
    c->p->recv(c);
  }
}
```

For truly parallel execution of code appropriate libraries are necessary. Some modifications to mapping algorithm may also be required.

# 8. CONCLUSION

Development of a pre-processor prototype and supporting tools for the markup language showed the following benefits of our approach.

Additional language constructions are not required to explain the meaning of the parallel algorithm. This is similar to approach based on object-oriented libraries STL [5], TBB [6], CCR [7], Boost [8], and others. However, the markup and pre-processing technique reduces the amount of manual coding.

More reliable protection against programming errors is provided. This feature is compatible with concurrent programming languages Go [9], Occam [10], Limbo [11], Erlang [12]. Static type checking in the implementation language helps to prevent incorrect connection of message source and message recipient. Semantic checking can also be implemented at the pre-processor level. For example, one can check the attainability of a state in the communication protocol for channels and the possibility to call a method for processes. Check can also be carried out during the program execution. If pair of processes does not perform a prescribed messaging protocol, calculations will stop.

Behavior of the marked program can be investigated in more detail by means of problem-oriented debugger [13]. The mapping algorithm can add code to provide information to the debugger.

The performance prediction of a parallel program is also possible. Discrete event simulation library can easily replace standard execution mechanism.

The markup language is a means of skeleton programming and code reuse [14,15]. One can design a universal skeleton for programs with similar control flow and adapt it to specific applications. The adaptation is made by the changing of message variables and handlers. This technique can be used for programming multi-core and many-core systems [16, 17].

The markup can be presented in form of a visual language. Translation from graphic elements to markup obviously follows from the samples in Appendix B. In contrast to the classical visual languages [13,18], a visual representation of a markup language is an additional means of preliminary design and documentation. Diagrams are used to present a single objects as in [19].

The markup language defines parallel execution with sequential code. This technique is used in incremental parallelization. A number of well-known [20, 21] and experimental [22, 23] tools for defining iterative or recursive parallelism are based on markup. We adopted the same method for a process-channel programming model.

Markup language can be applied to different languages and parallel implementation of a run-time library. It is compatible with the modern technologies [24,25] used in industrial software development.

## ACKNOWLEDGMENTS

My thanks go to Samara State Aerospace University (SSAU) for its continuous support to this research. The language run-time library was tested in the university supercomputer 'Sergey Korolev' with ongoing assistance of maintenance team. I should also mention a great contribution from my students and collaborators.

This work was supported by the Ministry of Education and Science of the Russian Federation in the framework of the implementation of the Program of increasing the competitiveness of SSAU among the world's leading scientific and educational centers over the period from 2013 till 2020.

## APPENDIX A. The code structure

```
module =   {base-language|user-block} module-scheme {base-language|user-block}.
user-block = user-prefix base-language user-postfix.
module-scheme =  scheme-prefix { channel | process } scheme-postfix.
```

## APPENDIX B. The syntax of channels and processes

```
channel = '~' ident [params] ['=' state {';' state}] '.'.
state = ['+'] ident [ ('?'|'!')  [rules] ].
rules = rule { '|' rule }.
rule = ident { ',' ident } '->' ident.
process = '*' ident [params] ['=' ((ports [';' actions]) | actions) ] '.'.
ports = port {';' port}.
port  = ident ':' ident ('?'|'!')[(rules ['|' '->' ident])|( '->' ident)].
actions = action {';' action}.
action  = ['+'] [ident ':'] disjunction ['->' ([ident] '|' ident) | ident].
disjunction = conjunction { '|' conjunction}.
conjunction = call {'&' call}.
call = ident '(' [args] ')'.
args = ident ('?'|'!') ident {',' ident ('?'|'!') ident }.
params = '<' ident {',' ident} '>'.
```

## APPENDIX C. The optional visual presentation for channels and processes

*Channel vertices*

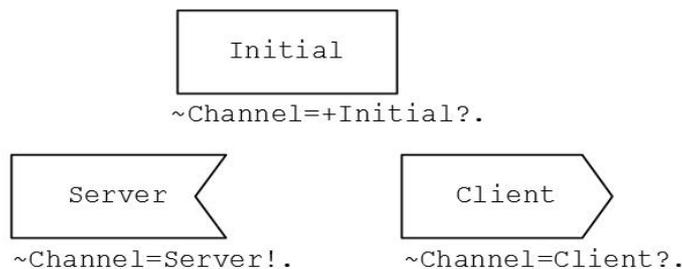

*Channel edge*

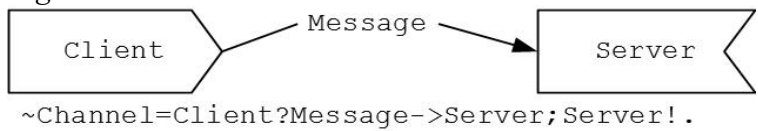

Note: the edge that denotes a message can join vertices of any type (`Initial`, `Server`, `Client`), but the Initial vertex should be the only one in the channel graph.

*Process vertices*

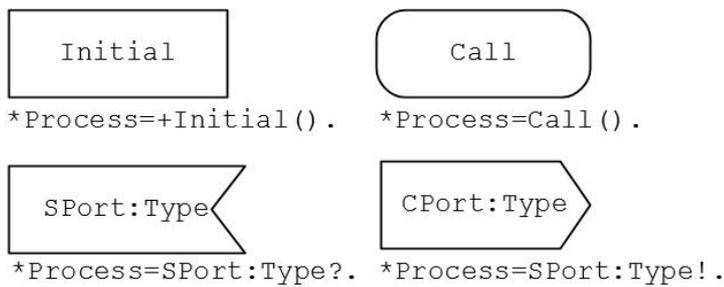

*Process data flow edges*

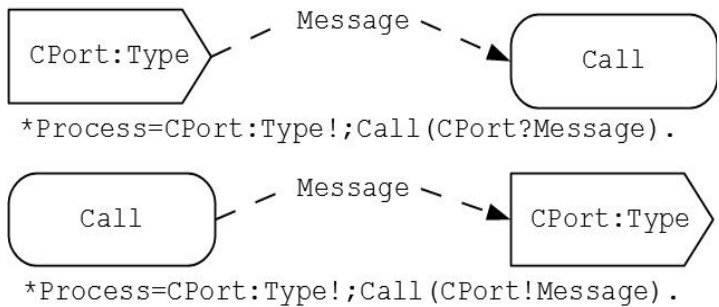

*Process control flow edges*

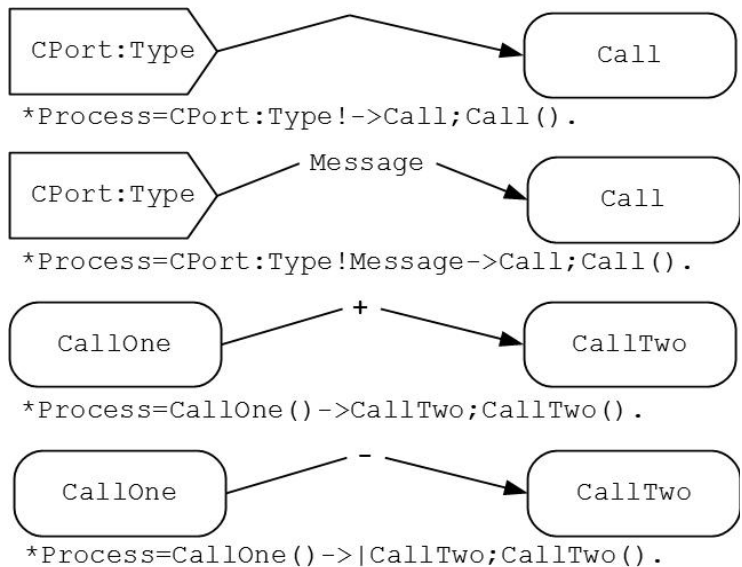

*Process mixed edge*

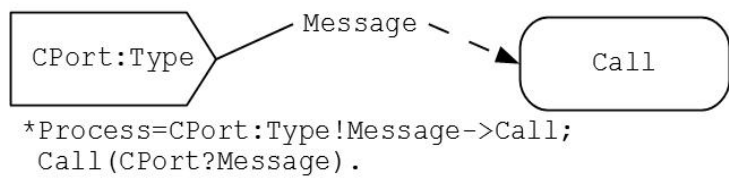

Note: the vertex that denotes a server port (`SPort:Type`) can be used instead of the client port vertex (`CPort:Type`) in any given pattern; the `Initial` vertex can also be used instead of the `Call` (and also `CallOne`, `CallTwo`) vertex, but the `Initial` vertex should be the only one in the process graph.

————————————————